\documentclass[a4papper,11pt,feynmf]{article}
\usepackage{amssymb}
\usepackage{graphicx}
\usepackage{ifpdf}
\begin{document}
\def\ctr#1{\hfil $\,\,\,#1\,\,\,$ \hfil}
\def\tstrut{\vrule height 2.7ex depth 1.0ex width 0pt}
\def\mystrut{\vrule height 3.7ex depth 1.6ex width 0pt}
\def \inparg{\leftskip = 40pt\rightskip = 40pt}
\def \outparg{\leftskip = 0 pt\rightskip = 0pt}
\def\lf{16\pi^2}
\def\beqn{\begin{eqnarray}}
\def\eeqn{\end{eqnarray}}
\def\sy{supersymmetry}
\def\sic{supersymmetric}
\def\sa{supergravity}
\def\ssm{supersymmetric standard model}
\def\sm{standard model}
\def\ssb{spontaneous symmetry breaking}
\def\smgroup{$SU_3\otimes\ SU_2\otimes\ U_1$}
\def\app{{Acta Phys.\ Pol.\ }{\bf B}}
\def\anp{Ann.\ Phys.\ }
\def\cmp{Comm.\ Math.\ Phys.\ }
\def\fortphys{{Fort.\ Phys.\ }{\bf A}}
\def\ijmpa{{Int.\ J.\ Mod.\ Phys.\ }{\bf A}}
\def\jetp{JETP\ }
\def\jetpl{JETP Lett.\ }
\def\jmp{J.\ Math.\ Phys.\ }
\def\mpla{{Mod.\ Phys.\ Lett.\ }{\bf A}}
\def\nc{Nuovo Cimento\ }
\def\npb{{Nucl.\ Phys.\ }{\bf B}}
\def\physrep{Phys.\ Reports\ }
\def\plb{{Phys.\ Lett.\ }{\bf B}}
\def\pnas{Proc.\ Natl.\ Acad.\ Sci.\ (U.S.)\ }
\def\pr{Phys.\ Rev.\ }
\def\prd{{Phys.\ Rev.\ }{\bf D}}
\def\prl{Phys.\ Rev.\ Lett.\ }
\def\ptp{Prog.\ Th.\ Phys.\ }
\def\sjnp{Sov.\ J.\ Nucl.\ Phys.\ }
\def\tmp{Theor.\ Math.\ Phys.\ }
\def\pw{Part.\ World\ }
\def\zpc{Z.\ Phys.\ {\bf C}}
\def\pa{\partial}

\vskip .3in {\textbf{\Large{ The role  of field redefinition on   renormalisability   of a general  $N=\frac{1}{2}$ supersymmetric  gauge theories}}}
\medskip
\vskip .3in \centerline{\bf   A.~F.~Kord $^{1,2}$,
M.~Haddadi Moghaddam $^{1}$, N.~Ghasempour $^{1}$  }
\bigskip
{\small{\it {$1$: Department of Physics, Hakim Sabzevari
University (HSU), P.O.Box 397, Sabzevar, Iran

 $2$: School of Particles and Accelerators, Institute for Research  in Fundamental Sciences(IPM),P.O. Box 19395-5531, Tehran, Iran }}}


\small{\it\center{E-mail: afarzaneh@hsu.ac.ir}} \vskip .3in

\begin{abstract}
We investigate   some issues on renormalisability  of non-anticommutative supersymmetric gauge theory related to field redefinitions. We  study one loop corrections to $N=\frac{1}{2}$ supersymmetric $SU(N)\times U(1)$ gauge theory coupled to chiral matter in component formalism, and show the procedure  which has been introduced  for renormalisation is problematic because some terms which are needed for the renormalisability of theory are missed from the Lagrangian. In order to prove the  theory is renormalisable, we   redefine  the  gaugino and the auxiliary fields($\lambda, \bar F$), which result in a modified form of the  Lagrangian in the component formalism.  Then, we show the modified   Lagrangian  has  extra terms  which are necessary for renormalisability   of
non-anticommutative  supersymmetric gauge field theories. Finally  we prove  $N = \frac{1}{2}$ supersymmetric gauge theory is renormalisable up to one loop corrections using  standard method of renormalisation; besides, it is shown  the effective action is gauge invariant.  
 \end{abstract}

\section{Introduction}
In recent years, deformed quantum field theories have received more attention due to their natural appearance in string theory. One type of deformation is space-time deformation in which the commutators of the space time coordinates become non-zero, which results in non-commutative field theories. The  non-commutativity leads  to  a nontrivial product of fields which is called the $\star$-product.  Another type of deformation is noncommutativity in the Grassmann coordinates $\theta^{\alpha}$,  leaving the anticommutators of $\bar{\theta}^{\dot{\alpha}}$ unchanged.  It has been indicated  that this superspace geometry can occur in string theory in the presence of a graviphoton background~\cite{a1,a2}. Theories defined on non-anticommutative (NAC) superspace have been studied extensively during last ten years  ~\cite{a1,a22,a23,a24,a3}. It is straightforward to construct a field theory over  NAC superspace in terms of superfields with the star-product where the  Lagrangian is deformed from the original theory by the non-anticommutativity
parameter $\{\theta^{\alpha}, \theta^{\beta}\}=C^{\alpha \beta}$ where $C$ is a nonzero constant.

During the last ten years, the renormalisability    of  the
NAC field theories has been   the subject of numerous research. NAC field theories are not
power-counting renormalisable; however, it has been shown that
they could be made renormalisable if some additional terms are
added to the Lagrangian in order to absorb divergences to all
orders~\cite{a4}-\cite{a9}. The issues of  renormalisability  of  NAC  versions of the Wess-Zumino model and supersymmetric gauge theories have been studied. The renormalisability of NAC versions of the Wess-Zumino model has been
discussed ~\cite{a4,a5}, with explicit computations up to two loops~\cite{a6}.
The renormalisability of supersymmetric gauge field theories has been discussed in WZ gauge~\cite{a7,a8}. Drawing on  the component approach,
 authors~\cite{a10,a11}  have provided the most complete results for the one-loop quantum corrections of the deformed
component theory. Working in components in the WZ gauge, they have argued that in order to restore gauge invariance, it is necessary to define  one-loop divergent field redefinitions  of the gaugino field($\lambda$) and the auxiliary field($\bar{F}$) (in the case of matter fields). It has been manifested,  the one-loop divergences(1PI), with the relevant  diagrams containing only $C$-deformed vertex figures, cannot be cancelled by the  Lagrangian since they contain contributions  which do not appear in the original Lagrangian; in other words, the theory is nonrenormalisable.  Their results imply that there are problems with the assumption of gauge invariance,   which is required to rule out some classes of divergent structure in the NAC theory. In  their findings, one can see that even at one loop divergent non-gauge-invariant terms are generated. In order to remove the
non-gauge-invariant terms and restore gauge invariance at one loop they have introduced  one loop divergent field redefinitions.
They realised that by adding new deformation-parameter-dependent terms to the theory, the one-loop effective action can be renormalisable. However, these kinds of divergent field redefinitions are problematic because there is no
theoretical justification or interpretation for the field redefinition, as mentioned by the authors ~\cite{a10,a11}.

On the other hand, the authors in~\cite{a13,a133,a134} started from the superspace formalism and  discussed renormalisability and  supergauge  invariance. They have    argued that suitable deformations of the classical actions are necessary in order to achieve renormalisability at one loop level~\cite{a13}.
  Using the background field method,  they have computed   the effective action. They have  proved that  divergent field redefinitions are not required and the original  effective action is not only gauge but also supergauge invariant up to one loop corrections. An important feature of their work is that although they obtain the effective action without any difficulty in the superspace formalism, they have found that some new  terms have to be added to the superfield action due to the one loop divergent contribution for $C$ deformed section.

In this paper we investigate the renormalisability of $N
=\frac{1}{2}$ supersymmetric   gauge theory  coupled to  chiral  matter  and  propose  a modified  classical  action  which  is necessary in component formalism. First,  we briefly review NAC supersymmetric gauge theories and their Lagrangian in the component formalism and also the field redefinitions which are described  in refs~\cite{a1,a3}. Then we concentrate on  the one-loop corrections  of  three and four-point functions ( in the C-deformed sector) and show that anomalous  terms  appear in the 1PI functions which spoil the renormalisability of the theory. In order to absorb these anomalous terms and renormalise the  theory  we  suggest   a new kind of  field redefinition which results in a new form of the  Lagrangian  in  the component formalism (though  the form of the  Lagrangian remains unchanged in the superspace formalism). Then, we  investigate its effects on  the renormalisability of the  theory. We  shall prove   $N = \frac{1}{2}$ supersymmetric gauge theory  is renormalisable at one loop level, using the standard method of renormalisation without any need for  divergent field redefinitions. Our method confirms  that the   effective action is  gauge invariant which is consistent with  superspace formalism ~\cite{a13}.

  Working in the component case, we initially encounter some very important issues such as the field redefinition of the gaugino field $\lambda$ and auxiliary field $\bar F$ that Seiberg and other authors have introduced at the beginning of their extension of the standard theory. With these redefinitions some terms have been effectively removed from the Lagrangian; we reveal the necessity of restoring these hidden terms by new generalized redefinitions based on ~\cite{a1,a3}. Secondly, the effective action in the component case violates gauge invariance owing to some unusual terms--the so-called $Y$ terms.  Nonetheless, we confirm a  number of results in  both of the works in   ~\cite{a11, a13} relating to renormalisability of the theory and preservation of the algebraic structure of the star product.

\section{$N = 1/2$ supersymmetric $U(N)$  gauge theory Lagrangian}
In this section we review  the classical form of the  $N =\frac{1}
{2}$  supersymmetric   gauge theory Lagrangian. The $N = 1/2$ supersymmetric   gauge theory Lagrangian
  was first introduced in Ref.~\cite{a1,a3}  for the gauge group $U(N)$. However, as it was noted
in Refs. ~\cite{a10,a11}, at the quantum level the $U(N)$ gauge invariance cannot be retained
since the $SU(N)$ and $U(1)$ gauge couplings renormalise differently; and they have  obliged to
consider a modified $N =\frac{1}{2}$ invariant theory with the gauge group $SU(N)\times U(1)$.

The $U(N)$ gauge invariant supersymmetric Lagrangian for NAC superspace formalism is as follows:
\begin{eqnarray}
L&=&\int
d^2\theta d^2\bar{\theta}\ \bar{{\Phi}}\star e^{V}\star \Phi \nonumber\\&&+ \frac{1}{16kg^2}\Big(\int d^2\theta tr W^{\alpha }\star W_{\alpha}+ \int d^2\bar{\theta}\overline{W}_{\dot{\alpha}}\star \overline{W}^{\dot{\alpha}}\Big),
\end{eqnarray}
where  $W_{\alpha}$ and $\overline{W}_{\dot{\alpha}}$ are chiral and antichiral field strengths.  $V$, $\Phi$ and $\bar\Phi$ are vector, chiral and anti chiral superfield respectively.

When one discusses  the non-anticommutative theory, he or she  starts with the superspace
formalism. In the superspace gauge transformation, the gauge parameter is a
superfield. When  one  rewrites it  into the component formalism, it is necessary   to
impose the Wess Zumino gauge. Using this gauge fixing, one obtains  the component
gauge transformation, which is smaller than the original superspace gauge
transformation.

In Wess Zumino gauge, the vector superfield $V$ is presented as:
 \begin{eqnarray}
 V(y,\theta,\bar{\theta})=-(\theta \sigma^{\mu}\bar{\theta})v_{\mu}(y)+i\theta\theta\bar{\theta}\bar{\lambda}(y)-i\bar{\theta}\bar{\theta}\theta\lambda(y)+\frac{1}{2}\theta\theta\bar{\theta}\bar{\theta}(D-i\partial_{\mu}v^{\mu})(y),
 \end{eqnarray}
where we write $V=V^AR^A$ with $R^A$ being the generators of the gauge group $U(N)$. Performing a gauge transformation on the vector superfield in the Wess-Zumino gauge results in the gauge transformations of the component fields. They   are given by:
\begin{eqnarray}
\delta^{*}_{\varphi}A_\mu &=&-2\partial_{\mu}\varphi +i[\varphi , A_{\mu}],
\\
\delta^{*}_{\varphi}\lambda_{\alpha} &=&i[\varphi , \lambda_{\alpha}]+\frac{1}{4}(\varepsilon C \sigma^{\mu})_{\alpha\dot{\alpha}}\lbrace -2\partial_{\mu}\varphi +\bar{\lambda}^{\dot{\alpha}}\rbrace ,
\\
\delta^{*}_{\varphi}\bar{\lambda}_{\dot{\alpha}} &=&i[\varphi , \bar{\lambda}_{\dot{\alpha}}],
\\
\delta^{*}_{\varphi}D&=&i[\varphi ,D],
\end{eqnarray}

where $\varphi$  is the gauge transformation parameter. These  gauge transformations are not canonical because the transformation of $ \lambda $
 depends on the deformation parameter C. In order to obtain the canonical form of the gauge transformations, the authors~\cite{a1,a3}  proposed  the following   $ \lambda $ redefinition:
 \begin{eqnarray}
{\lambda}^{'}_{\alpha}&=&\lambda_{\alpha} -\frac{1}{4}(\varepsilon C \sigma^{\mu})_{\alpha\dot{\alpha}}\lbrace A_{\mu},\bar{\lambda}^{\dot{\alpha}}\rbrace,
\end{eqnarray}
so that  its canonical  gauge transformation  is given by:
\begin{eqnarray}
\Longrightarrow \delta^{*}_{\varphi}{\lambda}^{'}_{\alpha} & = & i[\varphi ,{\lambda}^{'}_{\alpha}].
\end{eqnarray}
Then, the vector superfield is redefined as:
\begin{eqnarray}
V^A(y,\theta,\bar{\theta})&=&-(\theta\sigma^{\mu}\bar{\theta})A_\mu^A(y)+i\theta\theta\bar{\theta}\bar{\lambda}^A(y)
-i\bar{\theta}\bar{\theta}\theta\Big({\lambda}_{\alpha}^{'A}+\frac{1}{4}d^{ABC} C^{\mu\nu} \sigma_{\nu \alpha\dot\alpha} A_{\mu}^B\bar{\lambda}^{\dot\alpha C}\Big)\nonumber\\&&+\frac{1}{2}\theta\theta\bar{\theta}\bar{\theta}(D^A-i\partial^{\mu}A_{\mu}^A)(y)
\end{eqnarray}

Gauge transformations of the chiral and antichiral superfields have been studied in ~\cite{a1,a3}.
The chiral and anti chiral superfields are written as:
\begin{eqnarray}
{\Phi}(y,{\theta})&=&{\phi}(y)+\sqrt{2}{\theta}{\psi}(y)+{\theta}{\theta}{F}(y)\\
\bar{\Phi}(\bar{y},\bar{\theta})&=&\bar{\phi}(\bar y)+\sqrt{2}\bar{\theta}\bar{\psi}(\bar y)+\bar{\theta}\bar{\theta}\bar{F}(\bar y)
\end{eqnarray}
 In order to have canonical gauge transformations of the component fields, the $\bar{F}$ component field should be redefined as:
\begin{eqnarray}
{\bar{F}}^{'}(\bar{y})&=&\bar{F}(\bar{y})-iC^{\mu\nu}\partial_{\mu}(\bar{\phi}A_{\nu})(\bar{y})+\frac{1}{4}C^{\mu\nu}\bar{\phi}A_{\mu}A_{\nu}(\bar{y})
\end{eqnarray}
Then
\begin{eqnarray}
\Longrightarrow \delta^{*}_{\varphi}{\bar{F}^{'}}&=&-i{\bar{F}^{'}}\varphi
\end{eqnarray}
Thus, the antichiral superfield is given by\cite{a3}:
\begin{eqnarray}
\bar{\Phi}(\bar{y},\bar{\theta})&=&\bar{\phi}+\sqrt{2}\bar{\theta}\bar{\psi}+\bar{\theta}\bar{\theta}\Big({\bar{F}^{'}}+iC^{\mu\nu}\partial_{\mu}(\bar{\phi}A_{\nu})-\frac{1}{4}C^{\mu\nu}\bar{\phi}A_{\mu}A_{\nu}\Big)
\end{eqnarray}
Using  the above field  redefinitions  and rescaling $V$ and $C^{\alpha \beta}$, the authors~\cite{a3,a11}  have suggested  an $N = \frac{1}{2}$
 supersymmetric U(N) gauge theory action coupled to chiral matter. It is given by:

\begin{eqnarray}
S&=&\int
d^4x\Big[Tr\{-\frac{1}{2}F^{\mu\nu}F_{\mu\nu}-2i{\bar\lambda}{\bar\sigma}^\mu{(D_\mu{\lambda^{'}})}+
D^2\}\nonumber\\&&-2igC^{\mu\nu}Tr\{F_{\mu\nu}{\bar\lambda}{\bar\lambda}\}
+g^2\mid C\mid^2Tr\{({\bar\lambda}{\bar\lambda})^2\}\nonumber\\&&+\Big\{
{\bar{F}^{'}}F-i\bar{\psi}\bar{\sigma}^{\mu}(D_{\mu}\psi)-D_{\mu}\bar{\phi}D^{\mu}\phi+g\bar{\phi}D\phi+i\sqrt{2}g(\bar{\phi}{\lambda^{'}}\psi-\bar{\psi}\bar{\lambda}\phi)\nonumber\\&&+igC^{\mu\upsilon}\bar{\phi}F_{\mu\upsilon}F+{\sqrt{2}}gC^{\mu\nu}D_{\mu}\bar{\phi}\bar{\lambda}\bar{\sigma}_{\nu}\psi+\frac{|C|^{2}}{4}g^2\bar{\phi}\bar{\lambda}\bar{\lambda}F
\nonumber\\&& +(\phi \rightarrow\tilde{\phi}, \psi \rightarrow \tilde{\psi}, F \rightarrow \tilde{F} , C^{\mu,\nu} \rightarrow -C^{\mu,\nu})
\Big\}\Big],
\end{eqnarray}
where in order to   ensure anomaly cancellation,  a multiplet $ \{\phi,\psi, F\}$ transforming according to the fundamental representation
and, a multiplet $\{\tilde{\phi},\tilde{\psi},\tilde{F}\}$ transforming according
to its conjugate are  included in the Lagrangian. Moreover, there are

 \begin{eqnarray}
F_{\mu\nu}&=&\partial_\mu A_\nu-\partial_\nu A_\mu+ig[A_\mu , A_\nu],\nonumber\\
D_\mu\lambda^{'} &=&\partial_\mu\lambda^{'}+ig[A_\mu , \lambda^{'} ],  \nonumber\\ D_\mu\phi&=&\partial_\mu\phi+igA_\mu,
\end{eqnarray}
and
\begin{eqnarray}
A_\mu=A_\mu^AR^A,\ \lambda^{'}=\lambda^{'A}R^A,\ D=D^AR^A.
\end{eqnarray}
Corresponding to any index $a$ for SU(N), we introduce the index
$A = (0, a)$. Thus,  $A$ runs from 0 to $N^2 - 1$, with $R^A$
being the group matrices for U(N) in the fundamental
representation. These satisfy
\begin{eqnarray}
[R^A , R^B]&=&if^{ABC}R^C, \ \{R^A , R^B\}=d^{ABC}R^C,
\end{eqnarray}
where $f^{ABC}$ is completely antisymmetric, $f^{abc}$ is the same
as  $SU(N)$ and  $f^{0bc}=0$, while $d^{ABC}$ is completely
symmetric; $d^{abc}$ is the same as  $SU(N)$, $d^{0bc}=\sqrt{
2/N}\delta^{bc},\  d^{00c}=0$ and  $d^{000} = \sqrt{2/N}$. In
particular, $R^0 = \sqrt{ \frac{1}{2N}} 1$, and
\begin{eqnarray}
Tr\{R^AR^B\}=\frac{1}{2}\delta^{AB}
\end{eqnarray}

$C^{\mu\nu}$ is related to the non-anti-commutativity parameter
$C^{\alpha\beta}$ by:
\begin{eqnarray}
C^{\mu\nu}&=&C^{\alpha\beta}\epsilon_{\beta\gamma}\sigma_\alpha^{\mu\nu\
\gamma},
\end{eqnarray}
and
\begin{eqnarray}
\mid C\mid^2&=&C^{\mu\nu}C_{\mu\nu}.
\end{eqnarray}
Besides,  our conventions are consistent  with ref~\cite{a1}.

It is easy to show  there are  some  extra terms in the action when one  uses the original definition of the gaugino field $(\lambda)$ and the auxiliary field $(\bar{F})$ instead of  the  field redefinitions ~($\lambda^{'}$) and ($\bar{F}^{'}$). Thus, the  problem of renormalisability  of  the theory is solved by these extra terms as it is shown in next section.
\subsection{$N =\frac{ 1}{2}$ supersymmetric $SU(N)\times U(1)$ action }

To ensure renormalisability it is necessary to decompose $U(N)$ into $SU(N) \times  U(1)$ because the $U(N)$ gauge symmetry is not preserved under renormalisation. In fact, the
two gauge coupling constants renormalise differently~\cite{a11, a13}. To obtain a renormalisable theory one must introduce different couplings for the $SU(N)$ and $U(1)$ parts of the gauge group and then the U(N) gauge-invariance is lost.
Therefore,  the authors of \cite{a11} have suggested an $N=\frac{1}{2}$ supersymmetric  $SU(N)\times U(1)$ gauge theory coupled to chiral matter. Following their work, the action is given by:

\begin{eqnarray}
S&=&\int d^4x\Big[-\frac{1}{4}F^{\mu\nu A
}F_{\mu\nu}^A-i{\bar\lambda}^A{\bar\sigma}^\mu{(D_\mu\lambda^{'})}^A+\frac{1}{2}
D^AD^A\nonumber\\&&-\frac{1}{2}i\gamma^{ABC}d^{ABC}C^{\mu\nu}F_{\mu\nu}^A{\bar\lambda}^B{\bar\lambda}^C
+\frac{1}{8} \mid
C\mid^2 d^{abe}d^{cde}g^2({\bar\lambda}^a{\bar\lambda}^b)({\bar\lambda}^c{\bar\lambda}^d)\nonumber\\&&
+\frac{1}{4N} \mid
C\mid^2 (\frac{g^2}{g_0})^2({\bar\lambda}^a{\bar\lambda}^a)({\bar\lambda}^b{\bar\lambda}^b)\nonumber\\&&
+\{\bar F^{'}F-i\bar\psi\bar\sigma^\mu D_\mu\psi-D^\mu\bar\phi D_\mu\phi
+\bar\phi\hat D\phi+i\sqrt{2}(\bar\phi\hat{\lambda}^{'}\psi-\bar\psi\bar{\hat\lambda}\phi)\nonumber\\&&
+\sqrt{2}C^{\mu\nu}D_\mu\bar\phi\bar{\hat\lambda}\bar\sigma_\nu\psi+iC^{\mu\nu}\bar\phi\hat{F}_{\mu\nu}F
+\frac{1}{4}\mid C\mid^2 \bar\phi \bar{\hat\lambda}^B\bar{\hat\lambda}^C F    \nonumber\\&& +(\phi \rightarrow\tilde{\phi}, \psi \rightarrow \tilde{\psi}, F \rightarrow \tilde{F} , C^{\mu\nu} \rightarrow -C^{\mu\nu})
\}\Big],
\end{eqnarray}
 where $\hat{A}_\mu$ is  defined by
\begin{eqnarray}
\hat{A}_\mu=\hat{A}_\mu^AR^A=g{A}_\mu^aR^a+g_0{A}_\mu^0R^0,
\end{eqnarray}
with similar definitions for $\hat\lambda^{'}, \hat D$ and $\hat{F}_{\mu\nu},$ and
\begin{eqnarray}
D_\mu\phi=(\partial_\mu+i\hat{A}_\mu)\phi.
\end{eqnarray}
$\gamma^{ABC}$ is given by:
\begin{eqnarray}
\gamma^{abc}=g,\ \gamma^{a0b}=\gamma^{ab0}=\gamma^{000}=g_0,\ \gamma^{0ab}=\frac{g^2}{g_0}
\end{eqnarray}

Eq.~(22) reduces to the original U(N) Lagrangian Eq.~(15) derived from nonanticommuting
superspace upon setting $g_0 = g$. 

In order to investigate the renormalisability of the theory, one needs to compute the one-loop one-particle-irreducible(1PI) graph  contributions. The
one-loop graph corrections of $N = 1$ part of the theory are not affected by
$C$-deformation. So the anomalous dimensions and gauge $\beta$-functions are as for $N = 1$. The 1PI graph corrections contributing to the new terms (those containing C) are calculated in ref~\cite{a11} in the component formalism and we present them in the Appendix (We note that  the one loop divergent contributions for the C deformed sector have also  been computed   using  the background field method in  the superspace approach  in ref ~\cite{a13}). However, it is easy to see the component version of the theory could not be  renormalisable  because some anomaly terms which are proportional to ${Y^{\mu\nu}}=C^{\mu\rho}g_{\rho\lambda}({{\bar\sigma}^{\lambda\nu}})$ appear in 1PI corrections, but there are no similar terms in the Lagrangian; therefore, these terms spoil the renormalisability of the theory. Moreover, the existence of these terms violates the gauge invariance of the effective action (albeit this issue does not happen in the effective action according to superspace formalism). In order to solve these issues and    renormalise  the theory in component formalism, the authors of ref~\cite{a11} have proposed a procedure of two steps. Firstly, they have modified the action and  added new terms to the action. Their modified action  is given by:

\begin{eqnarray}
S&=&\int d^4x\Big[-\frac{1}{4}F^{\mu\nu A
}F_{\mu\nu}^A-i{\bar\lambda}^A{\bar\sigma}^\mu{(D_\mu\lambda^{'})}^A+\frac{1}{2}
D^AD^A\nonumber\\&&-\frac{1}{2}i\gamma^{ABC}d^{ABC}C^{\mu\nu}F_{\mu\nu}^A{\bar\lambda}^B{\bar\lambda}^C
+\frac{1}{8} \mid
C\mid^2 d^{abe}d^{cde}g^2({\bar\lambda}^a{\bar\lambda}^b)({\bar\lambda}^c{\bar\lambda}^d)\nonumber\\&&
+\frac{1}{4N} \mid
C\mid^2 (\frac{g^2}{g_0})^2({\bar\lambda}^a{\bar\lambda}^a)({\bar\lambda}^b{\bar\lambda}^b)\nonumber\\&&
+\frac{1}{N}\vartheta_1g_0^2C^2(\bar\lambda^a\bar\lambda^a)(\bar\lambda^0\bar\lambda^0)
+\{\bar F^{'}F-i\bar\psi\bar\sigma^\mu D_\mu\psi-D^\mu\bar\phi D_\mu\phi
+\bar\phi\hat D\phi+i\sqrt{2}(\bar\phi\hat{\lambda}^{'}\psi-\bar\psi\bar{\hat\lambda}\phi)\nonumber\\&&
+\sqrt{2}C^{\mu\nu}D_\mu\bar\phi\bar{\hat\lambda}\bar\sigma_\nu\psi+iC^{\mu\nu}\bar\phi\hat{F}_{\mu\nu}F
+\frac{1}{4}\mid C\mid^2 \bar\phi \bar{\hat\lambda}^B\bar{\hat\lambda}^C F  \nonumber\\&&
  -\vartheta_2 C^{\mu\nu}g\Big(\sqrt{2}D_\mu\bar\phi\bar\lambda^aR^a\bar\sigma_\nu\psi+\sqrt{2}\bar\phi\bar\lambda^aR^a\bar\sigma_\nu D_\mu\psi+i\bar\phi F_{\mu\nu}^aR^aF \Big)
  \nonumber\\&& +(\phi \rightarrow\tilde{\phi}, \psi \rightarrow \tilde{\psi}, F \rightarrow \tilde{F} , C^{\mu\nu} \rightarrow -C^{\mu\nu})
\}\Big],
\end{eqnarray}

where $\vartheta_1$ and $\vartheta_2$ are  constants. These new  terms ( those are proportional to $\vartheta_1$ and $\vartheta_2$ constants)    are   separately invariant under $N = 1/2$ supersymmetry and must be included to obtain a renormalisable Lagrangian. However, these terms are not obtained from  the original superspace action. In this case a similar feature also appeared in ~\cite{a13,a133,a134} where the superspace  action had to be modified  in order  to get a renormalised theory. In their procedure,  the renormalised couplings $\vartheta_1$ and $\vartheta_2$ have been  set to zero for calculational simplicity. In other words, they have not contributed to 1PI corrections. Secondly, they have  introduced  divergent field redefinitions or, to put it another way, added  non-linear terms to the bare action at the end of their calculations. This scenario has been used in several papers~\cite{a12}.
 However, the  scenario   is problematic because  of  divergent field redefinitions which have no theoretical justification. The another problem is that  the action is changed. Moreover, there is  disagreement with  the superfield formalism   where divergent field redefinitions are not needed~\cite{a13}.

In next section  we  introduce  field redefinitions which  lead to  a  different   classical action  in the component formalism, then discuss the  renormalisability of the theory, and prove the theory is  renormalisable   up to one loop corrections. Besides, we indicate   divergent field redefinitions are not needed which is in agreement with the  superfield formalism.

\section{ Generalized Wess Zumino gauge, Field redefinitions and Renormalisation}
In this section we generalise Wess Zumino gauge, and introduce field redefinitions which modify the  $N=\frac{1}{2}$ supersymmetric  gauge theory action. Then we prove the modified action is renormalisable.

Discussing  the non-anticommutative theory, one starts from the superspace formalism. In the superspace gauge transformation, the gauge parameter is a superfield. In order to obtain the $N=\frac{1}{2}$ supersymmetric  gauge theory Lagrangian in the component formalism, one should  impose the Wess Zumino gauge. Using  this gauge fixing, the component gauge transformation is obtained which is smaller than the original superspace gauge transformation.
In order to obtain the canonical forms of the gauge transformations, Seiberg proposed to take the following Wess Zumino gauge \cite{a1}.
\begin{eqnarray}
V^A(y,\theta,\bar{\theta})&=&-(\theta\sigma^{\mu}\bar{\theta})A_\mu^A(y)+i\theta\theta\bar{\theta}\bar{\lambda}^A(y)
-i\bar{\theta}\bar{\theta}\theta\Big({\lambda}_{\alpha}^A+\frac{1}{4}d^{ABC} C^{\mu\nu} \sigma_{\nu \alpha\dot\alpha} A_{\mu}^B\bar{\lambda}^{\dot\alpha C}\Big)+ \nonumber\\&&\frac{1}{2}\theta\theta\bar{\theta}\bar{\theta}(D^A-i\partial^{\mu}A_{\mu}^A)(y).
\end{eqnarray}
What we do is to impose the generalised version of this
Wess Zumino gauge fixing in the form:

\begin{eqnarray}
V = \cdots -i\bar{\theta}\bar{\theta}\theta\Big({\lambda}_{\alpha}^A+\frac{1}{4}d^{ABC} C^{\mu\nu} \sigma_{\nu \alpha\dot\alpha} A_{\mu}^B\bar{\lambda}^{\dot\alpha C}(1+\kappa^{ABC})
\Big)
+ \cdots
\end{eqnarray}
For $\kappa^{ABC}=0$, it reduces to the Seiberg's case. For $\kappa^{ABC}=-1$, it reproduces the form of $\lambda$ before the redefinition
appearing in \cite{a1,a3}. We postulate that the difference between the two gauge fixing is related by a certain superspace gauge transformation. Therefore, we believe that the parameter $\kappa^{ABC}$ simply corresponds to a  choice of gauge fixing. In the explicit calculation, we will see that $\kappa^{ABC}$ is renormalised. It means that we change the gauge fixing condition during the renormalisation.
In the usual renormalisation method, we keep the gauge fixing condition. In this sense, our method does not look very natural conceptually
although it does not necessarily mean that our computation is problematic. If we would like to keep the same Wess Zumino gauge,
we can go back to the original gauge fixing after the 1-loop computation by putting the renormalised coupling $\kappa^{ABC}_{\rm ren} = 0$ by hand
in the renormalised Lagrangian.

In the same way we generalise  the anti-chiral superfield as follow:
\begin{eqnarray}
\bar{\Phi}(\bar{y},\bar{\theta})&=&\bar{\phi}+\sqrt{2}\bar{\theta}\bar{\psi}+\bar{\theta}\bar{\theta}\Big({\bar{F}}+iC^{\mu\nu}\partial_{\mu}(\bar{\phi}A_{\nu})-\frac{1}{4}C^{\mu\nu}\bar{\phi}A_{\mu}A_{\nu}\nonumber\\&&+iC^{\mu\nu}\kappa_AR^A \bar\phi\partial_\mu A_\nu^A-\frac{i}{8}\varepsilon_A gR^Af^{ABC}C^{\mu\nu}\bar\phi A_\mu^B A_\nu^ C\nonumber\\&&
+\frac{1}{8}h_{ABC}\mid C\mid^2g_Bg_C d^{ABC}R^A\bar\phi \bar{\lambda}^B\bar{\lambda}^C\Big)
\end{eqnarray}
For $h_{ABC}=\varepsilon_A=\kappa_A= 0$, it reduces to the Eq.~(14).

  Using the generalised vector and anti-chiral superfield, Eqs.~(28,29) result in field redefinitions $\lambda^{'}$ and $\bar F^{'}$ which are:

\begin{eqnarray}
&&\lambda^{'A} \longrightarrow\lambda^A-\frac{1}{4}\kappa^{ABC}
d^{ABC}C^{\mu\nu} A_\mu^B \sigma_{\nu} \bar\lambda^{C}\nonumber\\&&
\bar F^{'} \longrightarrow\bar F-iC^{\mu\nu}\kappa_AR^A \bar\phi\partial_\mu A_\nu^A+\frac{i}{8}\varepsilon_A gR^Af^{ABC}C^{\mu\nu}\bar\phi A_\mu^B A_\nu^ C\nonumber\\&&
+\frac{1}{8}h_{ABC}\mid C\mid^2g_Bg_C d^{ABC}R^A\bar\phi \bar{\lambda}^B\bar{\lambda}^C.
\end{eqnarray}

 These field redefinitions are similar to Eqs (7) and (12). According to the above equations, the  WZ gauge has been parameterised by  some new non zero parameters  $\kappa^{ABC}$, $\kappa_A$ and $\varepsilon_A$.
 The field redefinitions in eqs. (30) lead to  new  gauge and SUSY transformations which are not canonical because the
transformation of $\lambda$ and $\bar F$ depend on the NAC parameter $C$.
The gauge  transformations are given by:
\begin{eqnarray}
\delta_\varphi A_{\mu}^A& =&-2\partial_{\mu}\varphi^A-f^{ABC}\varphi^B
A_{\mu}^C,\nonumber\\
 \delta_\varphi{\bar\lambda}_{\dot\alpha}^A&=&-f^{ABC}\varphi^B
{\bar\lambda}_{\dot\alpha}^C,\nonumber\\
\delta_\varphi
{\lambda}_{\alpha}^A&=&-f^{ABC}\varphi^B-\frac{1}{2}\kappa^{ABC}
d^{ABC}C^{\mu\nu}\sigma_{\nu\alpha\dot\alpha}\partial_\mu\varphi^B\bar\lambda^{\dot\alpha
C }\nonumber\\
\delta_\varphi D^A&=&-f^{ABC}\varphi^B D^C,\nonumber\\
\delta_\varphi\phi&=&i\varphi\phi,\nonumber\\
 \delta_\varphi\bar\phi&=&-i\bar\phi\varphi,\nonumber\\
 \delta_\varphi\psi_\alpha&=&i\varphi\psi_\alpha,\nonumber\\
\delta_\varphi\bar\psi_{\dot\alpha}&=&-i\bar\psi_{\dot\alpha}\varphi\nonumber\\
\delta_\varphi F&=&i\varphi F,\nonumber\\
\delta_{\varphi}\bar{F}&=&-i\bar{F}\varphi
 -iC^{\mu\nu}\kappa_{A}R^{A}f^{ABC}\bar{\phi}\partial_{\mu}(\varphi^{B}A_{\nu}^{C})\nonumber\\&&
 +\frac{i}{2}\varepsilon_{A}gR^{A}f^{ABC}C^{\mu\nu}\bar{\phi}(\partial_{\mu}\varphi^{B})A_{\nu}^{C}\nonumber\\&&
 +\frac{i}{4}\varepsilon_{A}gR^{A}f^{ABC}f^{BDE}C^{\mu\nu}\bar{\phi} \varphi^{D}A_{\mu}^{E}A_{\nu}^{C}\nonumber\\&&
 +\frac{1}{4}h_{ABC}|C|^{2}g_{B}g_{C}d^{ABC}R^{A}f^{BDE}\bar{\phi}\varphi^{D}\bar{\lambda}^{E}\bar{\lambda}^{C}
\end{eqnarray}
 Moreover, the $N=\frac{1}{2}$ SUSY transformations are given by:
\begin{eqnarray}
&&\delta A_\mu^A=-i{\bar\lambda}^A{\bar\sigma}_\mu\epsilon ~,\nonumber\\&&
\delta\lambda_\alpha^A=i\epsilon_\alpha
D^A+{(\sigma^{\mu\nu}\epsilon)}_\alpha[F_{\mu\nu}^A+\frac{1}{2}iC_{\mu\nu}(\gamma^{ABC}+\frac{1}{2}\kappa^{ABC})d^{ABC}{\bar\lambda}^B{\bar\lambda}^C],\nonumber\\&&
\delta {\bar\lambda_{\dot\alpha}}^A=0~,\ \delta D^A=-\epsilon\sigma^\mu
D_\mu {\bar\lambda}^A,\nonumber\\&&
\delta\phi=\sqrt{2}\epsilon\psi,\ \delta\bar\phi=0,\nonumber\\&&
\delta\psi^\alpha=\sqrt{2}\epsilon^\alpha F,\ \delta\bar\psi_{\dot\alpha}=-i\sqrt{2}(D_\mu\bar\phi)(\epsilon\sigma^\mu)_{\dot\alpha},\nonumber\\&&
\delta F=0,\nonumber\\&&
 \delta\bar F=-iR^{A}\bar{\phi}\epsilon\lambda^{A} +\
 \sqrt{2}i\epsilon \sigma^{\mu}(\partial_{\mu}\bar\psi
 +igR^{A}A_{\mu}^{A}\bar\psi)\nonumber\\&&
 +i(1-\kappa_{A})C^{\mu\nu}R^{A}\bar{\phi}(\epsilon\sigma_{\nu}\partial_{\mu}\bar{\lambda}^{A}) -\frac{1}{4}(1-\varepsilon_{A})gR^{A}f^{ABC}C^{\mu\nu} \bar{\phi}A_{\mu}^{B}(\epsilon\sigma_{\nu}\bar{\lambda}^{C})
\end{eqnarray}
\subsection{The modified action}
 The field redefinitions  Eqs.~(30) lead to  a modified NAC action  in  component formalism. Therefore,  we should replace  Eq.~(22) by:
\begin{eqnarray}
S&=&\int d^4x\Big[-\frac{1}{4}F^{\mu\nu A
}F_{\mu\nu}^A-i{\bar\lambda}^A{\bar\sigma}^\mu{(D_\mu\lambda^{})}^A+\frac{1}{2}
D^AD^A\nonumber\\&&-\frac{1}{2}i\gamma^{ABC}d^{ABC}C^{\mu\nu}F_{\mu\nu}^A{\bar\lambda}^B{\bar\lambda}^C
+\frac{1}{8} \mid
C\mid^2 d^{abe}d^{cde}g^2({\bar\lambda}^a{\bar\lambda}^b)({\bar\lambda}^c{\bar\lambda}^d)\nonumber\\&&
+\frac{1}{4N} \mid
C\mid^2 (\frac{g^2}{g_0})^2({\bar\lambda}^a{\bar\lambda}^a)({\bar\lambda}^b{\bar\lambda}^b)
+\frac{1}{N}\vartheta_1g_0^2C^2(\bar\lambda^a\bar\lambda^a)(\bar\lambda^0\bar\lambda^0)
\nonumber\\&&
\frac{i}{16}
d^{ABC}\kappa^{BAC}C^{\mu\nu}(\partial_\mu A^A_\nu-\partial_\nu
A^A_\mu)\bar\lambda^B\bar\lambda^C\nonumber\\&&-\frac{i}{4}
g\kappa^{EDB}d^{BDE}f^{ACE} C^{\mu\nu} A_\mu^C A_\nu^D
\bar\lambda^A \bar\lambda^B\nonumber\\&&+\frac{i}{4}
\kappa^{BAC}d^{ABC}A_\mu^A(\partial_\nu
{\bar\lambda}^BY^{\mu\nu}{\bar\lambda}^C-{\bar\lambda}^BY^{\mu\nu}\partial_\nu{\bar\lambda}^C)\nonumber\\&&+\frac{i}{2}
g\kappa^{EDB}f^{ACE}d^{BDE} A_\mu^C A_\nu^D
{\bar\lambda}^A Y^{\mu\nu} {\bar\lambda}^B\nonumber\\&&
+\{\bar F^{}F-i\bar\psi\bar\sigma^\mu D_\mu\psi-D^\mu\bar\phi D_\mu\phi
+\bar\phi\hat D\phi+i\sqrt{2}(\bar\phi\hat{\lambda}^{}\psi-\bar\psi\bar{\hat\lambda}\phi)\nonumber\\&&
+\sqrt{2}C^{\mu\nu}D_\mu\bar\phi\bar{\hat\lambda}\bar\sigma_\nu\psi+iC^{\mu\nu}\bar\phi\hat{F}_{\mu\nu}F
+\frac{1}{4}\mid C\mid^2 \bar\phi \bar{\hat\lambda}^B\bar{\hat\lambda}^C F  \nonumber\\&&
  -\vartheta_2 C^{\mu\nu}g\Big(\sqrt{2}D_\mu\bar\phi\bar\lambda^aR^a\bar\sigma_\nu\psi+\sqrt{2}\bar\phi\bar\lambda^aR^a\bar\sigma_\nu D_\mu\psi+i\bar\phi F_{\mu\nu}^aR^aF \Big)\nonumber\\&&
+i\frac{\sqrt{2}}{4}g_A\kappa^{ABC}d^{ABC}R^AC^{\mu\nu}\bar\phi A_\mu^B\bar\lambda^C\bar\sigma_\nu\psi\nonumber\\&&
-iC^{\mu\nu}\kappa_A R^A\bar\phi\partial_\mu A_\nu^A F+\frac{i}{8}\varepsilon_A gR^Af^{ABC}C^{\mu\nu}\bar\phi A_\mu^B A_\nu^ C F\nonumber\\&&
+\frac{1}{8}h_{ABC}\mid C\mid^2g_Bg_C d^{ABC}R^A\bar\phi \bar{\lambda}^B\bar{\lambda}^C F
   \nonumber\\&& +(\phi \rightarrow\tilde{\phi}, \psi \rightarrow \tilde{\psi}, F \rightarrow \tilde{F} , C^{\mu\nu} \rightarrow -C^{\mu\nu})
\}\Big],
\end{eqnarray}
where $\kappa^{ABC}$, $\kappa_A$ and $\varepsilon_A$ are some constants. Besides,  because of the renormalisability of  NAC $SU(N)\times U(1)$ gauge theory, we  require choosing
\begin{eqnarray}
&&\kappa^{ABC}=\xi\gamma^{BAC}c^Ac^Bd^C,\ \kappa_A=(\zeta c_A+\eta(1-c_A))g_A\nonumber\\&&  \ \kappa_a=\zeta g,\ \kappa_0=\eta g_0,\ \
\varepsilon_A=\tau g_A c^A,
\end{eqnarray}
where $\xi,\ \zeta,\ \eta,\ \tau$ and $h_{ABC}$ are some coefficients. Moreover,   $h_{ABC}$   depend on indices $A, B ,C$. We  note that $h_{a0b}=h_{ab0}$. In addition, $g_a\equiv g$, $c^A = 1- \delta^{A0}$, and  $d^A = 1 + \delta ^{A0}$.
We also have
\begin{eqnarray}
&&({Y^{\mu\nu}})^{\dot\alpha\dot\beta}=\epsilon^{\dot\alpha\dot\theta}C^{\mu\rho}g_{\rho\lambda}({{\bar\sigma}^{\lambda\nu}})^{\dot\beta}_{{}{\dot\theta}}.
\end{eqnarray}
 According to the above equations, the  action  has been parameterised by  some new non zero parameters  $\kappa^{ABC}$, $\kappa_A$ and $\varepsilon_A$. Such parameters have not been  introduced in refs~\cite{a1, a11, a3} because  they have worked in spacial Wess Zumino gauge. However, the renormalisation procedure reveals   the necessity for these non-zero couplings. We have realised that if these new coefficients  are  zero, then some terms are hidden in the classical action, meanwhile divergent contributions due to Feynman diagrams produce them.



It is straightforward to show that Eq.~(33) is preserved under the  gauge, and SUSY transformations of Eqs.~(31,32).

\subsection{Renormalisation of the  $SU(N) \times U(1)$  modified  action}

The divergences in one-loop diagrams should be cancelled by the one-loop
divergences in bare action, obtained by replacing the fields and couplings in Eq.~(33) with bare
fields and couplings given by

\begin{eqnarray}
&&A^a_{B\mu}=Z_{A^a}^{\frac{1}{2}}A^a_{\mu},\ \       \ \ A^0_{B\mu}=Z_{A^0}^{\frac{1}{2}}A^0_{\mu} \nonumber\\&&   \lambda^a_B=Z_{\lambda^a}^{\frac{1}{2}}\lambda^a,  \ \
 \lambda^0_B=Z_{\lambda^0}^{\frac{1}{2}}\lambda^0, \ \
\phi_B=Z^{\frac{1}{2}}_{\phi}\phi, \ \ \psi_B=Z^{\frac{1}{2}}_{\psi}\psi, \ \  g_B=Z_gg\nonumber\\&&
 C^{\mu\nu}_B=Z_C C^{\mu\nu},\ \ \mid
C\mid^2_B=Z_{\mid C\mid^2}\mid C\mid^2\nonumber\\&&
\xi_B=Z_\xi \xi,\ \zeta_B=Z_\zeta \zeta,\ \eta_B=Z_\eta \eta,\nonumber\\&&
\tau_B=Z_\tau \tau ,\ h_{(ABC)_{Bare}}=Z_{h_{ABC}}h_{ABC}\nonumber\\&&
\vartheta_{1B}=Z_{\vartheta_1}\vartheta_1,\ \vartheta_{2B}=Z_{\vartheta_2}\vartheta_2.
\end{eqnarray}
In Eq.~(36) we have set the renormalised
couplings  $\xi, \zeta, \eta, \tau, h_{ABC}, \vartheta_1, \vartheta_2$    to zero for simplicity. The other renormalisation constants
start with tree-level values of $1$. Therefore, we have:
\begin{eqnarray}
&&
\xi_B=Z_\xi^{(1)},\ \zeta_B=Z_\zeta^{(1)},\ \eta_B=Z_\eta^{(1)},\nonumber\\&&
\tau_B=Z_\tau^{(1)}, ,\ h_{(ABC)_{ Bare}}=Z_{h_{ABC}}^{(1)},\nonumber\\&&
\vartheta_{1B}=Z_{\vartheta_1}^{(1)},\ \vartheta_{2B}=Z_{\vartheta_2}^{(1)}
\end{eqnarray}

The $C$-independent one-loop corrections are cancelled by the one-loop  divergences in the $C$-independent part of the bare  action. Thus,
the renormalisation constants
for the fields and for the gauge couplings $g, g_0$ are the same as in the ordinary $N = 1$
supersymmetric theory~\cite{a7}, and up to one loop corrections they  are given  by~\cite{a0}:
\begin{eqnarray}
&&Z_\lambda=1-2L(N+1),\ Z_{\lambda^0}=1-2L,\nonumber\\&&
Z_A=1+2L(N-1),\ Z_{A^0}=1-2L,\nonumber\\&&
Z_g=1+L(1-3N),\ Z_{g_0}=1+L,\nonumber\\&&
Z_\phi=1,\ Z_\psi=1-2L\hat C_2,
\end{eqnarray}
where (using dimensional regularisation with $d = 4 - \epsilon$ )$ L =\frac{g^2}{16\pi^2\epsilon}$ and
\begin{eqnarray}
\hat C_2=(N+\frac{1}{N}\Delta)
\end{eqnarray}
with
\begin{eqnarray}
\Delta=(\frac{g_0}{g})^2-1
\end{eqnarray}

Upon inserting Eq. (38) into Eq. (33) one could  obtain the one-loop contributions from $S_{Bare}$ as

\begin{eqnarray}
S_{Bare}^{(1)}&=&\int d^4x\Big((4NL+2L)igC^{\mu\nu}d^{abc}\partial_\mu A_\nu^a\bar\lambda^b\bar\lambda^c+4iLg_0C^{\mu\nu}d^{ab0}\partial_\mu A_\nu^a\bar\lambda^b\bar\lambda^0\nonumber\\&&+(8NL+2L)i\frac{g^2}{g_0}C^{\mu\nu}d^{0bc}\partial_\mu A_\nu^0\bar\lambda^b\bar\lambda^0+2iLg_0 C^{\mu\nu}d^{000}\partial_\mu A_\nu^0\bar\lambda^0\bar\lambda^0\nonumber\\&&-(3NL+L)ig^2C^{\mu\nu}d^{abe}f^{cde}A_\mu^cA_\nu^d\bar\lambda^a\bar\lambda^b\nonumber\\&&
-(2NL+2L)igg_0C^{\mu\nu}d^{0be}f^{cde}A_\mu^cA_\nu^d\bar\lambda^0\bar\lambda^b\nonumber\\&&
-(\frac{5}{4}NL+\frac{1}{4}L)g^2\mid C\mid^2d^{abe}d^{cde}(\bar\lambda^a\bar\lambda^b)(\bar\lambda^c\bar\lambda^d)\nonumber\\&&+(4\frac{g^2}{g_0^2}L+\frac{1}{2N}L)\mid C\mid^2(\bar\lambda^a\bar\lambda^a)(\bar\lambda^b\bar\lambda^b)\nonumber\\&&
-\frac{i}{2}Z_C^{(1)}\gamma^{ABC}d^{ABC}C^{\mu\nu}F_{\mu\nu}^A{\bar\lambda}^B{\bar\lambda}^C
\nonumber\\&&+Z_{\mid C\mid^2}^{(1)}\Big[\frac{1}{8} \mid
C\mid^2 d^{abe}d^{cde}g^2({\bar\lambda}^a{\bar\lambda}^b)({\bar\lambda}^c{\bar\lambda}^d)
+\frac{1}{4N} \mid
C\mid^2 (\frac{g^2}{g_0})^2({\bar\lambda}^a{\bar\lambda}^a)({\bar\lambda}^b{\bar\lambda}^b)\Big]\nonumber\\&&+\sqrt{2}(-L\hat C_2-4NL)gC^{\mu\nu}\partial_\mu\bar\phi\bar\lambda^aR^a\bar\sigma_\nu\psi
+\sqrt{2}(-L\hat C_2)g_0C^{\mu\nu}\partial_\mu\bar\phi\bar\lambda^0R^0\bar\sigma_\nu\psi\nonumber\\&&
+i\frac{\sqrt{2}}{2}(6NL+L\hat C_2)g^2C^{\mu\nu} \bar\phi A_\mu^b \bar\lambda^ad^{abC}R^C\bar\sigma_\nu \psi\nonumber\\&&
+i\sqrt{2}(2NL+L\hat C_2)gg_0C^{\mu\nu} \bar\phi A_\mu^b \bar\lambda^0R^0 R^b\bar\sigma_\nu \psi\nonumber\\&&
+i\sqrt{2}(4NL+L\hat C_2)gg_0C^{\mu\nu} \bar\phi A_\mu^0 \bar\lambda^aR^a R^0\bar\sigma_\nu \psi\nonumber\\&&
+i\sqrt{2}(L\hat C_2)(g_0)^2C^{\mu\nu} \bar\phi A_\mu^0 \bar\lambda^0( R^0)^2\bar\sigma_\nu \psi\nonumber\\&&
+i(-4NL)C^{\mu\nu}gR^a\bar\phi\partial_\mu A_\nu^aF+i(4NL)g^2C^{\mu\nu}\bar\phi R^a f^{abc } A_\mu^b  A_\nu^cF\nonumber\\&&
+(-NL)g^2\mid C\mid^2 d^{Abc}R^A\bar\phi \bar{\lambda}^b\bar{\lambda}^c F
+(-NL)gg_0\mid C\mid^2 d^{a0c}R^a\bar\phi \bar{\lambda}^0\bar{\lambda}^c F\nonumber\\&&+\sqrt{2}Z_C^{(1)}C^{\mu\nu}D_\mu\bar\phi\bar{\hat\lambda}\bar\sigma_\nu\psi
+Z_C^{(1)}iC^{\mu\nu}\bar\phi\hat{F}_{\mu\nu}F
+\frac{1}{4}Z_{\mid C\mid^2}^{(1)}\mid C\mid^2 \bar\phi \bar{\hat\lambda}^B\bar{\hat\lambda}^C F\nonumber\\&&
+\frac{i}{8}(Z_\xi^{(1)}) g d^{abc}C^{\mu\nu}
 \partial_{\mu} A_\nu^a
\bar\lambda^b \bar\lambda^{c}
- \frac{i}{8}(Z_\xi^{(1)}) g^2  f^{cde} d^{abe} C^{\mu\nu} A_{\mu}^c A_\nu^d
\bar\lambda^a \bar\lambda^{b}
\nonumber\\&&
+ \frac{i}{4}(Z_\xi^{(1)})\xi g  d^{abc}
(\partial_{\mu} \bar\lambda^b Y^{\mu\nu}  \bar\lambda^{c}
- \bar\lambda^b Y^{\mu\nu} \partial_{\mu} \bar\lambda^{c} )
A_\nu^a\nonumber\\&&
-\frac{i}{4}(Z_\xi^{(1)}) g^2  f^{abe} d^{cde}  A_{\mu}^c A_\nu^d
\bar\lambda^a Y^{\mu\nu} \bar\lambda^{b}
+\frac{i}{4}(Z_\xi^{(1)}) g_0 d^{ab0}C^{\mu\nu}
\partial_{\mu} A_\nu^a
\bar\lambda^b \bar\lambda^{0}\nonumber\\&&
+ \frac{i}{2}(Z_\xi^{(1)}) g_0 d^{ab0}
(\partial_{\mu} \bar\lambda^b Y^{\mu\nu}  \bar\lambda^{0}
- \bar\lambda^b Y^{\mu\nu} \partial_{\mu} \bar\lambda^{0} )
A_\nu^a\nonumber \\&&
- \frac{i}{2}(Z_\xi^{(1)}) g_0 g  f^{cde} d^{0be} C^{\mu\nu} A_{\mu}^c A_\nu^d
\bar\lambda^0 \bar\lambda^{b}\nonumber\\&&
+\frac{i\sqrt{2}}{4}(Z_\xi^{(1)})g^2C^{\mu\nu} \bar\phi A_\mu^b \bar\lambda^cd^{abc}R^a\bar\sigma_\nu \psi+\frac{i\sqrt{2}}{2}(Z_\xi^{(1)})gg_0C^{\mu\nu} \bar\phi A_\mu^b \bar\lambda^0d^{ab0}R^a\bar\sigma_\nu \psi\nonumber\\&&
-i(Z_\zeta^{(1)})C^{\mu\nu} g R^a \bar\phi\partial_\mu A_\nu^a F
-i(Z_\eta^{(1)})C^{\mu\nu} g_0 R^0 \bar\phi\partial_\mu A_\nu^0 F\nonumber\\&&
+\frac{i}{8}(Z_\tau^{(1)}) g^2C^{\mu\nu}R^af^{abc}\bar\phi A_\mu^b A_\nu^c F+\frac{1}{4}(Z_{h_{ABC}}^{(1)})\mid C\mid^2 \bar\phi \bar{\hat\lambda}^B\bar{\hat\lambda}^C F\nonumber\\&&
-\sqrt{2}(Z_{\vartheta_2}^{(1)})gC^{\mu\nu}g \partial_\mu\bar\phi\bar\lambda^aR^a\bar\sigma_\nu\psi\nonumber\\&&
-\sqrt{2}(Z_{\vartheta_2}^{(1)})gC^{\mu\nu}g \bar\phi\bar\lambda^aR^a\bar\sigma_\nu\partial_\mu\psi
-2i(Z_{\vartheta_2}^{(1)})gC^{\mu\nu}R^a\bar\phi\partial_\mu A_\nu^a F\nonumber\\&&
+i(Z_{\vartheta_2}^{(1)})g^2C^{\mu\nu}\bar\phi R^af^{abc}A_\mu^b A_\nu^c F
+\frac{1}{N}(Z_{\vartheta_1}^{(1)})g_0^2\mid C\mid^2(\bar\lambda^a\bar\lambda^a)(\bar\lambda^0\bar\lambda^0)
\Big)
\end{eqnarray}

The results $\Gamma_{i-1PI}^{(1)},\ i=1,...8$ for the one loop divergences from the 1PI graphs coming from the $C$ dependent part of the $N=1/2$  supersymmetric gauge theory coupled to matter  are given in Appendix A\cite{a11}.  We find that with
\begin{eqnarray}
&&Z_C^{(1)}=Z_{\mid C\mid^2}^{(1)}=0,\ Z_\xi^{(1)}=-2NL,\ Z_{\vartheta_2}^{(1)}=-NL\\&&
 Z_\zeta^{(1)}=-(5N+2\hat C_2)L,\ Z_\eta^{(1)}=2\hat C_2L,\ Z_\tau^{(1)}=-(18N+8\hat C_2)L\\&&
Z_{h_{abc}}^{(1)}=\frac{-37N+32\hat C_2}{8}L,\ Z_{h_{ab0}}^{(1)}=-2(N-\hat C_2)L\\&&
Z_{h_{0bc}}^{(1)}=\frac{3}{4}NL,\ Z_{h_{000}}^{(1)}=2L\hat C_2,\ Z_{\vartheta_1}^{(1)}=-3NL,
\end{eqnarray}
they can be canceled  by Eq.~(41). In fact,
we have
\begin{eqnarray}
&&S_{Bare}^{(1)}+\sum_{i=1}^{8} \Gamma_{i-1PI}^{(1)}=finite,
\end{eqnarray}

 Our results indicate  the theory is renormalisable in the usual procedure~\cite{a00}  without using  one-loop divergent field redefinitions of the gaugino and the auxiliary fields($\lambda, \bar F$).  However, it is necessary to include the terms involving $ \vartheta_1, \vartheta_2$ in Eq.~(33) since further divergent configurations arise at one-loop which are $ N = 1/2$ supersymmetric. These terms are not in the original formulation of the theory  though they are independently $N = 1/2$ supersymmetric.   Therefore, one  should  modify the classical superspace  Lagrangian Eq.~(1) because these terms  are not obtained from the original  superfield action Eq.~(1). This point   is consistent with results \cite{a13,a133,a134}. They have modified  the classical action Eq.~(1) in order to have  one-loop renormalisable theory.

 In our work, we modify   the classical  action of Eq~(26) in order to make  the theory be  renormalisable and gauge invariant. In the modified action Eq~(33),   there are extra terms  which were absent in ~\cite{a11}. In our model,  we have  new renormalised couplings ($\xi, \zeta, \eta, \tau, h_{ABC}$) which start with tree-level values of zero for simplicity. In order to  renormalise the theory  we use the field  redefinitions of  the gaugino and the auxiliary fields($\lambda, \bar F$) in component formalism, but the classical superspace action is not modified. In fact we use a different Wess Zumino gauge in compare with that  used in refs~\cite{a1,a3}. Moreover,  we obtain the same divergent contributions $Z_{\vartheta_1}^{(1)}, Z_{\vartheta_2}^{(1)}$ as those in Ref~\cite{a11} which is a good  check of our results.

In our work we use the following  gaugino  field redefinition
\begin{eqnarray}
\lambda^{'A}\longrightarrow\lambda^A-\frac{1}{4}\kappa^{ABC}
d^{ABC}C^{\mu\nu} A_\mu^B \sigma_{\nu} \bar\lambda^{C}
\end{eqnarray}
or, we have:
 \begin{eqnarray}
 \delta\lambda^A=\lambda -\lambda^{'}= \frac{1}{4}\kappa^{ABC}
d^{ABC}C^{\mu\nu} A_\mu^B \sigma_{\nu} \bar\lambda^{C}= \frac{1}{4} \xi\gamma^{BAC}c^Ac^Bd^C d^{ABC}C^{\mu\nu} A_\mu^B \sigma_{\nu} \bar\lambda^{C}
 \end{eqnarray}

 After the renormalisation, we have
\begin{eqnarray}
\delta\lambda^A=\frac{1}{2}NL\gamma^{BAC}c^Ac^Bd^C
d^{ABC}C^{\mu\nu} A_\mu^B \sigma_{\nu} \bar\lambda^{C},
\end{eqnarray}
   which is similar to the  nonlinear  gaugino  field redefinition  introduced in   ref~\cite{a11}.  Now we can give an interpretation of the nonlinear  gaugino  field redefinition in  \cite{a11}. They have worked in Seiberg's Wess Zumino gauge ( $\kappa ^{ABC}=0$  in Seiberg's parametrisation but this choice is not preserved in the renormalisation)   which is not a suitable convention for the renormalisation,  then they have been forced  to  use the nonlinear field redefinition. However, in order to renormalise the theory we use a generalized Wess Zumino gauge where our redefinition is associated with some parameters, so we do not need to use the nonlinear field redefinition.  The same comparison between our auxiliary field ($ \bar F$) redefinition    the divergent  auxiliary field redefinition  of ref~\cite{a11} can be used in order to interpret  the nonlinear auxiliary field redefinition. In fact, they have used  the nonlinear  field redefinitions   to absorb unusual divergent contributions  which  are produced by 1PI graphs, and   have  found that variation of $\lambda$ and $\bar F$  result in  a change in the action; besides,  adding these divergent contributions  to the classical action the the theory is  renormalisable   up to one loop corrections.   In this sense, we would like to conclude that the divergent field redefinitions used in \cite{a11} were actually correct, although their interpretation was not clearly written.

We have obtained $Z_{C^2} = |Z_C|^2=1$ which means  the non-anticommutative structure is preserved by the renormalisation
despite the fact that the modified action (Eq.~41) has an explicit dependence on the NAC parameter.
Therefore, the star product does not get deformed by quantum corrections which is  consistent with ref~\cite{a11}  for the case of the component formalism and  ref~\cite{a13} for the case of the superspace formalism.

The authors of Ref~\cite{a13} have studied  one-loop quantum properties of the deformed  superspace theory and showed that the one-loop effective action could be renormalised if one modifies the NAC action in superspace formalism and our work should be compared to their work.
 Generally, we confirm their work although some of the details may differ. Working in superspace,  in a
background field approach, they have shown  that new divergences were present which  cannot be renormalised away. In order to make the theory be renormalisable
they have modified the classical action from the start by adding new terms which allow for the cancellation of all the divergent terms at one loop. We have taken the same approach by adding new terms to the Lagrangian from start in the  component formalism.   We prove that subtraction of one–
loop divergences does not require  nonlinear field redefinitions which is consistent with \cite{a13}, and also  the discussion  is cleaner. The important point however was the check that indeed, even in the presence of NAC, the effective action is gauge invariant and therefore the safety of going
to WZ gauge is ensured.

In this work we assume  the renormalised
couplings $\xi, \zeta, \eta, \tau, \vartheta_1, \vartheta_2$ and $h_{ABC}$ are set to zero. These assumptions simplify our calculations; in other words, we do not  consider contributions from terms which are proportional to these couplings   to the loop divergences. However, in a future extended  work  we  will consider non-zero values for these renormalised couplings, and calculate their contributions to quantum corrections . In our previous paper\cite{a14}, we have computed one-loop corrections which come from extra terms  in $N=1/2$ supersymmetric pure gauge theory. Moreover,  Our results are consistent with ref~\cite{a13} which used the superspace formalism. In both cases, in order to  obtain a renormalisable Lagrangian it is vital to add  some new terms to  the original Lagrangian. 
\section{Conclusion}
We have investigated the renormalisability of a general $N=\frac{1}{2}$ supersymmetric $SU(N)\times U(1) $ gauge theory coupled to chiral matter    at one loop order. We have proved the theory is renormalisable up one loop order using the standard method  of  renormalisation by adding some extra terms   which are generated by field redefinitions  of the gaugino and the auxiliary fields($\lambda, \bar F$), and   some new terms which are put by hand to the original  component Lagrangian. Moreover,  we  have shown the effective action is gauge invariant up to  one-loop corrections.


We have indicated  there is no need to employ divergent  redefinitions of $\lambda$ and $\bar F$ .  We have used the $N=\frac{1}{2}$ gauge group $SU(N)\times U(1)$ because of    the requirements of gauge
 invariance and renormalisability. As discussed in \cite{a4} the non-anticommutative  $SU(N)$ gauge  theory  is not   well-defined, and the non-anticommutative  $U(N)$ gauge  theory  is not renormalisable\cite{a11,a13}.

  We have shown that  the problem of the renormalisability  of the non-anticommutative theory  in the component formalism can be solved by field redefinitions. One of advantages of the field redefinition method is that it does not change the original Lagrangian in the  superspace formalism; in other words, Eq~(1) is preserved under field redefinitions and the theory is renormalised.
We have proved that the complete divergent part of the effective action which come from $C$-deformed section is gauge invariant even though term by term these quantum corrections are not gauge invariant, and  also arrived at the conclusion that there is no need to renormalise the non-anticommutativity parameter $C$,    which is consistent with Ref~\cite{a13}.
\section{Acknowledgements}
It is a pleasure to thank F. Yagi  for helpful discussions. We thank Ian Jack and Mohammad Reza Amirian for their help.
\appendix

\section{Divergent contributions for 1PI graphs}
The divergent contributions to the effective action from the $C$ dependent diagrams with one gauge, two gaugino lines $(A_\mu\bar\lambda\bar\lambda)$ are given by:

\begin{eqnarray}
\Gamma_{1-1PI}^{(1)}&=&-(\frac{15}{4}NL+2L)igC^{\mu\nu}d^{abc}\partial_\mu A_\nu^a\bar\lambda^b\bar\lambda^c-4iLg_0C^{\mu\nu}d^{ab0}\partial_\mu A_\nu^a\bar\lambda^b\bar\lambda^0\nonumber\\&&
-(8NL+2L)i\frac{g^2}{g_0}C^{\mu\nu}d^{0bc}\partial_\mu A_\nu^0\bar\lambda^b\bar\lambda^0-2iLg_0 C^{\mu\nu}d^{000}\partial_\mu A_\nu^0\bar\lambda^0\bar\lambda^0\nonumber\\&&+\frac{i}{2}NLg_0C^{\mu\nu}d^{ab0}\partial_\mu A_\nu^a\bar\lambda^b\bar\lambda^0-NLg d^{abc}
 \bar\lambda^b Y^{\mu\nu} \partial_{\mu} \bar\lambda^{c}
A_\nu^a\nonumber\\&&-2iNLg_0d^{ab0}
 \bar\lambda^b Y^{\mu\nu} \partial_{\mu} \bar\lambda^{0}
\end{eqnarray}

The divergent contributions to the effective action from the $C$ dependent diagrams with two gauge and two gaugino lines $(A_\mu A_\nu\bar\lambda\bar\lambda)$ are:
\begin{eqnarray}
\Gamma_{2-1PI}^{(1)}&=&(\frac{11}{4}NL+L)ig^2C^{\mu\nu}d^{abe}f^{cde} A_\mu^c  A_\nu^d\bar\lambda^a\bar\lambda^b\nonumber \\&&+\frac{1}{2}NLi g^2  f^{abe} d^{cde}  A_{\mu}^c A_\nu^d
\bar\lambda^a Y^{\mu\nu} \bar\lambda^{b}\nonumber \\&&
+(NL+2L)ig_0 g  f^{cde} d^{0be} C^{\mu\nu} A_{\mu}^c A_\nu^d
\bar\lambda^0 \bar\lambda^{b}.
\end{eqnarray}
The divergent contributions to the effective action from the $C$ dependent diagrams with four gaugino lines $(\bar\lambda\bar\lambda)^2$ are:
\begin{eqnarray}
\Gamma_{3-1PI}^{(1)}&=&(\frac{5}{4}NL+\frac{1}{4}L)g^2\mid C\mid^2d^{abe}d^{cde}(\bar\lambda^a\bar\lambda^b)(\bar\lambda^c\bar\lambda^d)\nonumber\\&&+(4\frac{g^2}{g_0^2}L+\frac{1}{2N}L)\mid C\mid^2(\bar\lambda^a\bar\lambda^a)(\bar\lambda^b\bar\lambda^b)\nonumber\\&&+3NLg_0^2\mid C\mid^2(\bar\lambda^a\bar\lambda^a)(\bar\lambda^0\bar\lambda^0
\end{eqnarray}

The divergent contributions to the effective action from the $C$ dependent diagrams with one gaugino, one scalar and one chiral fermion
line $(\bar\phi\bar\lambda\psi)$ are given by:
\begin{eqnarray}
\Gamma_{4-1PI}^{(1)}&=&\sqrt{2}(L\hat C_2+3NL)gC^{\mu\nu}\partial_\mu\bar\phi\bar\lambda^aR^a\bar\sigma_\nu\psi\nonumber\\&&
+\sqrt{2}(L\hat C_2)g_0C^{\mu\nu}\partial_\mu\bar\phi\bar\lambda^0R^0\bar\sigma_\nu\psi\nonumber\\&&
+\sqrt{2}(-NL)gC^{\mu\nu}\bar\phi\bar\lambda^aR^a\bar\sigma_\nu\partial_\mu\psi
\end{eqnarray}
 The divergent contributions to the effective action from the $C$ dependent diagrams with one gaugino, one scalar, one chiral fermion and
one gauge line $(A_\mu\bar\phi\bar\lambda\psi)$ are:
\begin{eqnarray}
\Gamma_{5-1PI}^{(1)}&=&i\sqrt{2}NLg^2C^{\mu\nu}A_\mu^b\bar\phi\bar\lambda^a\bar\sigma_\nu\psi[\frac{1}{2}d^{abc}R^c-2if^{abc}-3d^{abC}R^C]\nonumber\\&&
+i\sqrt{2}NLgg_0C^{\mu\nu}A_\mu^0\bar\phi\bar\lambda^a R^aR^0\bar\sigma_\nu\psi[-4]\nonumber\\&&
-i\sqrt{2}NLg^2C^{\mu\nu}A_\mu^a\bar\phi\bar\lambda^bR^aR^b\bar\sigma_\nu\psi[\hat C_2]\nonumber\\&&
-i\sqrt{2}NLgg_0C^{\mu\nu}A_\mu^0\bar\phi\bar\lambda^bR^0R^b\bar\sigma_\nu\psi[\hat C_2]\nonumber\\&&
-i\sqrt{2}NLgg_0C^{\mu\nu}A_\mu^a\bar\phi\bar\lambda^0R^aR^0\bar\sigma_\nu\psi[\hat C_2]\nonumber\\&&
-i\sqrt{2}NLg_0^2C^{\mu\nu}A_\mu^0\bar\phi\bar\lambda^0(R^0)^2\bar\sigma_\nu\psi[\hat C_2]
\end{eqnarray}

The divergent contributions to the effective action from the $C$ dependent    diagrams with one gauge, one scalar and one auxiliary line $(\bar\phi A_\mu F)$ are given by:
\begin{eqnarray}
\Gamma_{6-1PI}^{(1)}&=&i(-2L\hat C_2) gC^{\mu\nu}\bar\phi\partial_\mu A_\nu^a R^a F+i(-3NL)gC^{\mu\nu}\bar\phi\partial_\mu A_\nu^a R^a F\nonumber\\&&
+i(-2L\hat C_2)g_0C^{\mu\nu}\bar\phi\partial_\mu A_\nu^0 R^0 F
\end{eqnarray}

The divergent contributions to the effective action from the $C$ dependent diagrams with two gauge, one scalar and one auxiliary line are:
\begin{eqnarray}
\Gamma_{7-1PI}^{(1)}&=&i(L\hat C_2) g^2C^{\mu\nu}\bar\phi R^a f^{abc } A_\mu^b  A_\nu^cF\nonumber\\&&+i(-\frac{1}{4}NL)g^2C^{\mu\nu}\bar\phi R^a f^{abc } A_\mu^b  A_\nu^cF
\end{eqnarray}

The divergent contributions to the effective action from the $C$ dependent diagrams with two gaugino, one scalar and one auxiliary line $(\bar\phi\bar\lambda\bar\lambda F)$ are:
\begin{eqnarray}
\Gamma_{8-1PI}^{(1)}&=&(\frac{45}{8}NL)g^2\mid C\mid^2 d^{abc}R^a\bar\phi \bar{\lambda}^b\bar{\lambda}^c F\nonumber\\&&
+(-4L\hat C_2)g^2\mid C\mid^2 d^{abc}R^a\bar\phi \bar{\lambda}^b\bar{\lambda}^c F\nonumber\\&&
+(5NL)gg_0\mid C\mid^2 d^{0bc}R^c\bar\phi \bar{\lambda}^0\bar{\lambda}^b F\nonumber\\&&
+(-4L\hat C_2)gg_0\mid C\mid^2 d^{0bc}R^c\bar\phi \bar{\lambda}^0\bar{\lambda}^b F\nonumber\\&&
+(\frac{1}{4}NL)g^2\mid C\mid^2 d^{0bc}R^0\bar\phi \bar{\lambda}^b\bar{\lambda}^c F\nonumber\\&&
+(-2L\hat C_2)(g_0)^2\mid C\mid^2 d^{000}R^0\bar\phi \bar{\lambda}^0\bar{\lambda}^0 F
\end{eqnarray}

\end{document}